# AN EXPOSITION OF POSSIBILITY AND PROBABILITY

B. O'NEILL[**], *University of New South Wales*[**]

**Abstract**

This paper considers the notion of possible events which are insignificant in probabilistic analysis (i.e. events that have zero probability). The paper discusses the method of modal logic based on "possible worlds" and discusses a mathematical framework for the concepts of possibility, impossibility and certainty that are sometimes (incorrectly) thought to be defined with respect to probability. The relationship between possibility and probability is explored for general probability spaces and for refinements of these spaces conditional on other events, with particular focus on the properties of events having zero probability. We derive conditions under which possibility and significance diverge and conditions under which they can be reconciled as equivalent ideas within certain contexts. We also apply this analysis to discuss issues in possibility and probability in the multinomial model.

There are some strange creatures in probability theory and statistical science, and perhaps none is stranger than possible events having zero probability. These vicious beasts arise in problems involving continuous probability distributions, and are known for terrorising undergraduate students, though they have even devoured a professor or two, when the latter have wandered off-track without a proper guide. There are several challenges posed by these kinds of events. In the first place, their very existence cleaves probability away from possibility and impossibility, making it impossible to define the latter in probabilistic terms. In the second place, they leave many decision-makers uncertain as to how possibility and impossibility are relevant (or irrelevant) to their probabilistic conclusions.

The confusion engendered by these events stems from a flawed understanding of the modal relations of possibility and probability. This confusion stems in large part from the common practice of ascribing probabilistic definitions to possibility and necessity; in particular, to the common practice of speaking about an event as being possible if and only if it has positive probability. Clearly such a definition is incompatible with the existence of possible events having zero probability, and this generates immense confusion when such events are used in probability problems.

[*] Email address: ben.oneill@hotmail.com.
[**] School of Physical, Environmental and Mathematical Sciences, University of New South Wales, Australian Defence Force Academy, Canberra ACT 2600, Australia.

Although they are not equivalent, there is still substantial correspondence between the modal operation of possibility and the probability measure used in probability theory. In fact, when we are considering only a finite or countable set of possible events, we can avoid possible events with zero probability. However, once we consider events on sets of outcomes which are uncountably large, the divergence between the probability measures and the modal operator of possibility rears its head. In such cases we obtain events that are possible, but are also so implausible as to be irrelevant for the purposes of rational decision making when considered individually.

A proper understanding of the relationship between probability and possibility allows us to make statements regarding the possibility of events from our probabilistic beliefs and vice versa. Understanding the relationship in some precise mathematically defined sense, allows us to avoid confusions that arise from shorthand explanations of possibility in terms of probability. It is therefore useful to examine the relationship between these two concepts mathematically, by means of deriving an axiom system describing their interaction.

## 1. Modal logic, quantification, and "possible states of the world"

Modal logic is concerned with determining the logical rules governing assertions of necessity, possibility, deontological operators, temporal operators, and other logical operations having properties related to the quantifiers in standard predicate logic (see Lycan 1994 and Girle 2001). In particular, modal logic defines the logical operations of necessity and possibility and studies the rules of logic that apply to statements with these operations.[1] Unlike basic logic involving universally valid assertions of truth and falsity, modal logic allows statements of the form "It is *possible* that A", "It is *necessarily* true that B", "It *used to be* the case that C implied D", and so on.

Modal logic uses the rules of quantification in predicate logic to define various kinds of modal assertions and examine the logical rules pertaining to modal operators. This allows logicians to examine modal assertions which make some *qualified* assertion of truth, unlike the universally valid truth in standard predicate logic. As Putnam (1967) points out, introducing modal connectives, "...is not introducing new kinds of objects, but rather extending the kinds of things we can say about ordinary objects and sorts of objects" (p. 21).

[1] It is useful to note that the operations of necessity and possibility can actually be defined in terms of one another through logical negation. This means that modal logic only needs to deal formally with one of these operations, with the other being implicitly included. (Often both are explicitly included purely for elucidation.)

The validity of statements in modal logic is usually studied by reference to a set $\mathrm{W}$ of "states of the world" defined by reference to the modal operator under consideration. So, for example, modal statements involving possibility and necessity can be studied by considering the set of *possible* states of the world, whereas modal statements involving temporal operations can be studied by considering the set of states of the world *at particular times*. (These operators could be considered in conjunction by considering the set of *possible* states of the world *at particular times*.)

By considering modal statements about possibility and necessity in terms of possible states of the world, the validity of these statements can be assessed using standard logical methods with respect to each possible state $w \in \mathrm{W}$ and the overall truth of the modal assertion can then be resolved via quantification over all the states in $\mathrm{W}$. Something is *possibly true* if it is true in *some* possible state of the world, and it is *necessarily true* if it is true in *all* possible states of the world.

The task of establishing a philosophical definition for possibility falls into the domain of epistemology and the semantics of logic. Following Lewis (1986), quantificational methods for studying possibility and necessity have often been expressed using *possible world semantics* in which $\mathrm{W}$ is regarded as the set of "possible worlds". If we take this term "possible worlds" merely to be a shorthand way of expressing the fact that each *state* in $\mathrm{W}$ describes *the world* and is regarded as being *possible* then there is no problem with this. Lewis (1986) goes further than this, and has contended that the "possible worlds" used in modal logic are actual *existing* worlds which are each spatiotemporally and causally isolated from one another. Under this view (known as "modal realism") the notions of possibility and necessity refer to the actual existence of worlds separate from our own, with various properties of interest.

The present author regards this doctrine as both unjustified and unjustifiable. By their very definition, there can never be any evidence for the existence of worlds that are spatiotemporally and causally isolated from our own. Hence, the assertion that such worlds exist is arbitrary in the sense expounded in Peikoff (1993). Kripke (1980) has criticised the possible world semantics, and has rebuked the alleged existence of any "possible worlds" apart from the *actual* world. He correctly points out that this semantic view is, at best, merely a useful quantificational method for visualising the notion of possibility in another way (see pp. 43-45). (For more on quantification using possible worlds semantics see Kripke 1980, Divers 2002 and Girle 2003.) In our analysis we will therefore take a unitary view of existence under which there is only the *actual* world; the world that we live in. We will not need to incorporate any alleged distinction between logical and empirical possibility into our analysis and will proceed solely with a single notion of possibility as a personal epistemic concept, taken with respect to some particular individual.[2]

The field of modal logic involves the study of axioms for modal operators and their interaction. There are many different axiomatic systems in the literature, but the particular axiom system that most commends itself to modal reasoning is a task that is beyond the scope of this paper, and our present task is to look at the quantification over possible states of the world and see how this set of possible states interacts with the probability measure in probability theory —in particular, why do possible events with zero probability arise and what relevance (if any) do they have to our decisions?

## 2. Possibility and probability as epistemic concepts

Having accepted a unitary view of existence, in which the notion of possible worlds is merely shorthand for possible states of the *actual* world, we will proceed with our analysis by interpreting both possibility and probability as epistemic concepts. For probability, we take the epistemic interpretation that is used in the Bayesian paradigm used by de Finetti (1974) and subsequent Bayesian practitioners. We hold that the probability of an event refers to our subject's degree of confidence that the event will occur (or has occurred). Under this view, probability is a decision-making tool with no necessary ontological analogue in nature, so philosophical questions of determinism and indeterminism, and any other metaphysical bear-traps are neatly avoided.

[2] On the alleged dichotomy between logical and empirical possibility it is worth noting that this is part of the broader distinction between analytic and synthetic truths. The ontological notion of separately existing "possible worlds" forms the basis for the view that the properties of existents can be divided into those that are essential to the concept (i.e., those that must be present in all possible worlds) and those that are merely accidental (i.e., that could have be different in some other "possible world"). This distinction is present in various alleged dualisms, such as the modern analytic-synthetic dichotomy; for criticism of this view see Piekoff (1993).

This interpretation of probability is well-known to statisticians and has been discussed at length by Bayesian practitioners (see e.g. Bernardo and Smith 1994). In particular, it is well-known that a mathematical probability measure representing an epistemic degree of confidence can be derived from foundational arguments in decision theory using certain rationality desiderata (see Bernardo and Smith 1994, pp. 13-49; see also Fishburn 1986; Dupré and Tipler 2009). This leads to the use of a probability measure operating on a set of states of the world, and quantifying the degree of confidence in various events, each composed of states of the world (we will elaborate on the mathematics soon enough).

For the purposes of comparison with probability, the most coherent interpretation of possibility is also an epistemic interpretation, where the modal operator of possibility is taken with respect the knowledge of the same subject and refers to the consistency of an event with the knowledge of the subject. This is the interpretation adopted by Dubois (2006) in his discussion of possibility and probability (see pp. 47-49). This interpretation ensures that the concepts of possibility and probability can both be measured with respect to the knowledge of a particular subject, and are therefore comparable with one another. Moreover, it avoids metaphysical speculation on the true ontological state of the world and thereby sidesteps many nasty questions of metaphysics.

We will therefore follow the epistemic approach and take an assertion of possibility to mean that the possible outcome (i.e., the possible "state of the world") is *not known to be false* by the person of interest. By using this epistemic interpretation, we ensure that probability and possibility statements are both statements about the knowledge of a given person, and are therefore comparable with one another.

We reject the view, put forward by Hacking (1967), that possibility should be defined with respect to the aggregated knowledge of some "relevant community" —or, to put it another way, we hold that the "relevant community" is merely the individual to whom our possibility and probability statements apply (for discussion of this and other issues, see Teller 1972, Fetzer 1974 and DeRose 1991). This means that our statements of possibility and probability will both refer to the same individual observer, with a single set of knowledge, ensuring compatibility.

We also reject the suggestion by Hacking (1967) that possibility should be defined in terms of knowledge that includes information that could be obtained by "practicable investigation".[3] That is, when we talk of possibility (and its derivative concepts) we do so with respect only to that which is *actually known*, not that which *could* be known if such-and-such a thing were investigated (no matter how practicable the investigation might be). It is the present author's view that situations involving practically obtainable knowledge are better understood by being explicit about one's conditioning information, something which is already very easy to do within the framework of probability or possibility theory.

## 3. Representations of possibility and probability

There is a sizable literature analysing mathematical representations of possibility (this field is known as possibility theory), which is closely related to modal logic and fuzzy set theory. Overviews of this field can be found in Yager (ed) (1982), Kacprzyk and Orlovski (eds) (1987), Dubois and Prade (1988), Terano, Asai and Sugeno (1992), Zadeh and Kacprzyk (1992) and Dubois (2006).

At the outset of our discussion, it is important to note that possibility theory focuses on a wider representation of possibility than we are concerned with in a comparison of probability measures and the standard possibility operators that arise in modal logic. According to Dubois (2006), possibility theory can "...be viewed as a graded extension of modal logic where the dual notions of possibility and necessity already exist ... in an all-or-nothing format". The mathematics of possibility theory uses a "multi-valued logic" which accommodates a range of possibility values, beyond strict possibility and impossibility. The main purpose of this extension lies in the fact that the graded possibility values in possibility theory can be used to represent upper and lower probability bounds in the imprecise probability paradigm (see Walley 1991). Analysis of the relationship between mathematical possibility and probability representations, using multi-valued logics, can be found in Dubois and Prade (1993) and in Dubois (2006).

[3] Gibbs (1970) makes a similar proposal regarding the "available ways" of verifying that something is incompatible with one's knowledge.

In the present paper, we will focus on an all-or-nothing interpretation of possibility and will therefore confine ourselves to a level of generality that is narrower than that presented in possibility theory. We will accept (and use) the notion that there are degrees of belief about the truth of propositions. However, we reject the view that there are degrees of truth —that is, we accept the correspondence theory of truth and the law of the excluded middle, holding that all meaningful propositions are, in fact, either true or false, and not somewhere in between. With this in mind, we will not need to use multi-valued logic to express our notion of possibility, which can be represented as a simple indicator function.

This means that we will not need to make use of the full machinery of possibility theory in our exposition of possibility and necessity. We will instead consider a simple representation of possibility that is consistent with the present epistemological perspective and which elucidates the mathematical relationship between epistemic possibility and probability. Such a perspective only requires sufficient mathematical apparatus to separate all states of the world into those which have some property of interest and those that do not. (Possibility theory uses a sufficiently general apparatus to go beyond this dichotomy, if such is thought to be useful.) This will allow us to see exactly how little is needed to relate the two concepts and make sense of the notion of possible events with zero probability.

## 4. Probability measures and significant and almost sure events

Fishburn (1986) shows that basic rationality desiderata, completeness of preferences and various other measurement assumptions are sufficient to ensure the existence of quantitative beliefs that can be represented as a "probability measure" which is a normed set function meeting the axioms of probability in Kolmogorov (1933).

**DEFINITION 1 (Probability measures and probability spaces):** Let the **sample space** $\mathcal{S}$ be some set of **outcomes** and let $\mathfrak{G}$ be a sigma-field of **events** on $\mathcal{S}$ (i.e., groups of states of the world sharing some relevant similarity). A probability measure $\mathbb{P}: \mathfrak{G} \rightarrow \mathbb{R}$ is a mapping satisfying the following **axioms of probability**:

| | |
|---|---|
| **Non-negativity** | $\mathbb{P}(\mathcal{E}) \geq 0$ for all $\mathcal{E} \in \mathfrak{G}$, |
| **Norming** | $\mathbb{P}(\mathcal{S}) = 1$, |
| **Countable additivity** | $\mathbb{P}(\bigcup_{i=1}^{\infty} \mathcal{E}_i) = \sum_{i=1}^{\infty} \mathbb{P}(\mathcal{E}_i)$ for disjoint $\mathcal{E}_1, \mathcal{E}_2, \mathcal{E}_3, \ldots \in \mathfrak{G}$. |

The triple $(\mathcal{S}, \mathfrak{G}, \mathbb{P})$ is called a **probability space**.

**DEFINITION 2 (Insignificant and almost sure events):** An event $\mathcal{E} \in \mathfrak{G}$ is called an **insignificant** event if $\mathbb{P}(\mathcal{E}) = 0$ and is called an **almost sure** event if $\mathbb{P}(\mathcal{E}) = 1$.[4] If the event $\mathcal{E}$ is not insignificant so that $\mathbb{P}(\mathcal{E}) > 0$ then it is called a **significant** event.

Under the epistemic interpretation of probability, this measure is used to quantify the degree of confidence that the subject person has in the occurrence of any given event in the relevant sigma-field. Conditioning the analysis on the occurrence of any significant event yields a new conditional probability space as follows.

**DEFINITION 3 (Conditional probability space):** For any significant event $\mathcal{S}_0 \in \mathfrak{G}$ we let $\mathfrak{G}_0 \equiv \{\mathcal{E} \in \mathfrak{G}: \mathcal{E} \subseteq \mathcal{S}_0\}$ be a sub-sigma-field on $\mathcal{S}_0$ and we let $\mathbb{P}_0: \mathfrak{G}_0 \rightarrow \mathbb{R}$ be the conditional probability measure (conditional on $\mathcal{S}_0$) defined by:

$$\mathbb{P}_0(\mathcal{E}) \equiv \mathbb{P}(\mathcal{E})/\mathbb{P}(\mathcal{S}_0) \quad \text{for all } \mathcal{E} \in \mathfrak{G}_0.$$

The triple $(\mathcal{S}_0, \mathfrak{G}_0, \mathbb{P}_0)$ is also a probability space.

The reason that events with zero probability (insignificant events) are regarded as irrelevant to decision making in probability theory can be understood by looking at the effect of conditioning on an almost sure event. When we condition on an almost-sure event (so that we exclude an insignificant event from consideration) we obtain a correspondence between our original probabilistic beliefs and our new probabilistic beliefs, such that our analysis is simply reduced to the smaller space. To see this, we introduce the following simple result.

[4] The reason that these are referred to as "almost-sure" events, rather than "sure" events is a product of the terminology in measure theory. An event with probability one is "almost-sure" because there are some outcomes in the sample space that are not in the event.

**DEFINITION 4 (Reduction of probability spaces):** We will say that the probability space $(\mathcal{S}_0, \mathfrak{G}_0, \mathbb{P}_0)$ is a **reduction** of the probability space $(\mathcal{S}, \mathfrak{G}, \mathbb{P})$ if $\mathcal{S}_0 \subseteq \mathcal{S}$, $\mathfrak{G}_0 \subseteq \mathfrak{G}$ and $\mathbb{P}_0(\mathcal{E}) = \mathbb{P}(\mathcal{E})$ for all $\mathcal{E} \in \mathfrak{G}_0$.

**THEOREM 1:** The probability space $(\mathcal{S}_0, \mathfrak{G}_0, \mathbb{P}_0)$ is a reduction of the probability space $(\mathcal{S}, \mathfrak{G}, \mathbb{P})$ only if $\mathcal{S}_0$ is an almost sure event in $(\mathcal{S}, \mathfrak{G}, \mathbb{P})$ so that $\mathcal{S} - \mathcal{S}_0$ is an insignificant event in $(\mathcal{S}, \mathfrak{G}, \mathbb{P})$.

Theorem 1 shows us that conditioning on any almost sure event allows us to reduce our probability space by ignoring the corresponding insignificant event. This shows us that an event is insignificant if and only if it is *irrelevant* in the course of rational decision making —that is, if and only if removing it from the sample space does not affect the probabilistic beliefs for any events under consideration. Insignificant events can be ignored in our analysis, in the sense that, if we construct a new sample space excluding them from consideration then we obtain the same probabilities for all events under consideration.

This explains the reason why probability theory has no concern with different types of insignificant events. In particular, it has no concern with whether an insignificant event is possible or impossible. Such events simply have no relevance to rational inference or decision making in accordance with the rules of probability theory.

## 5. Possible and necessary events

We have already discussed our interpretation of possibility as a personal epistemic concept, taken with respect to the same individual and using the same conditioning knowledge as we have used for our probability space. From the point of view of a mathematical analysis of the interaction of possibility and probability, what matters is that these two ideas correspond, so that we avoid the incommensurate situation that would arise if probability referred to an epistemic concept pertaining to a certain person with certain conditioning knowledge, while possibility referred to, say, a metaphysical concept, or an epistemic concept pertaining to a different person (or people) or the same person with different conditioning knowledge. By insisting that possibility refers solely to the personal judgement of a single individual based on his actual knowledge we ensure that possibility and probability statements referring to particular events are comparable.

The notion of possibility applies to specific outcomes in a sample space and events in a sigma field of events. Since outcomes representing states of the world are either consistent or inconsistent with our actual personal knowledge of the nature of reality, it follows that all outcomes in a sample space can be classified as being either possible or impossible. We can therefore construct a *possibility space*.

**DEFINITION 5 (Possibility space):** Let the **possibility space** $\mathcal{W} \subseteq \mathcal{S}$ be the set of all possible states of the world. That is, we define $\mathcal{W}$ such that all outcomes $x \in \mathcal{W}$ are considered **possible** and all outcomes $x \in \mathcal{S} - \mathcal{W}$ are considered **impossible**.

From the possibility space we are able to derive the notion of possible, impossible, certain and uncertain events in a way that is consistent with the possibility and impossibility of particular outcomes. We are also able to define a possibility function which identifies each type of event. Mathematical modelling of the possibility of events on a sample space has been done by Zadeh (1978), leading to the concept of a possibility function. This kind of function has similar —though not identical— properties to a probability function.

**DEFINITION 6 (Possible events):** An event $\mathcal{E} \subseteq \mathcal{S}$ is **possible** if it contains at least one possible outcome and is **impossible** if it contains no possible outcomes. An event $\mathcal{E} \subseteq \mathcal{S}$ is **certain** if it contains every possible outcome and is **uncertain** if it does not contain every possible outcome.

Before proceeding to introduce the possibility function, it is worth noting that this transition from the possibility of particular outcomes to the possibility of events rests on the view that the disjunction of an arbitrary group of statements is false whenever the individual statements are all false. That is, if the statement "outcome $x$ is possible" is false for each value $x \in \mathcal{E}$ then the statement "event $\mathcal{E}$ is possible" must also be false.

While this position may seem intuitively reasonable, it rests on a form of consistency which is denied by some logical systems (this issue is related to the axiom of choice discussed in Cohen 1966). In the absence of this form of consistency, there is no logical inconsistency in asserting the impossibility of each particular outcome in an uncountable set, while simultaneously asserting the possibility of the event composed of those outcomes. This position would allow us to maintain the position that all insignificant events are impossible and all significant events are possible (see Zaman 1987). Notwithstanding this kind of potential objection, we will proceed on the basis of the standard possibility theory in which the disjunction of an arbitrary group of statements is false whenever all the individual statements are all false. This leads directly to the definition of possible and impossible events given above.

**DEFINITION 7 (Possibility function):** Given the possibility space $\mathcal{W} \subseteq \mathcal{S}$ we define the **possibility function** $\nabla: \mathfrak{G} \rightarrow \{0,1\}$ by $\nabla(\mathcal{E}) \equiv \mathbb{I}(\mathcal{E} \cap \mathcal{W} \neq \emptyset)$ for all $\mathcal{E} \in \mathfrak{G}$.[5]

**THEOREM 2 (Possibility of events):** Any event $\mathcal{E} \in \mathfrak{G}$ is possible if and only if $\nabla(\mathcal{E}) = 1$ and is impossible if and only if $\nabla(\mathcal{E}) = 0$. Moreover, any event $\mathcal{E} \in \mathfrak{G}$ is certain if and only if $\nabla(\mathcal{S} - \mathcal{E}) = 0$ and is uncertain if and only if $\nabla(\mathcal{S} - \mathcal{E}) = 1$. ■

The function $\nabla$ is a special case of the normalised possibility function defined in Dubois, Moral and Prade (1997) and in Dubois (2006). This function gives upper probabilities of events in imprecise probability under certain conditions (see Dubois and Prade 1992). These functions have been studied at least as far back as Zadeh (1978). In general, such functions can be shown to satisfy a similar set of properties to those applying to probability measures, with the countable-additivity property replaced by the property that the possibility of an arbitrary union of events is equal to the supremum of the possibility function over those events.

[5] The operator $\mathbb{I}$ is the "indicator function" for the argument statement.

It is again worth noting that more general rules for possibility functions have been designed to handle ranges of possibility values in the context of multi-valued logics (with varying interpretations of this idea). In fact, there are a number of possible mathematical formulations for propositional calculus with various philosophical interpretations (e.g. Shafer 1976, Zadeh 1978, Walley 1991; see Dubois and Prade 1993 for discussion).

From these properties, we have now established a suitable mathematical structure for the notion of possibility, interpreted, as we have stated, as a personal epistemic concept. In particular, given the notion of possible outcomes in the sample space, the derivative notions of possible, impossible, certain and uncertain events correspond to our general understanding of possibility and impossibility. We are now able to investigate the relationship between possibility and probability.

## 6. Relationship between probability and possibility

So far we have established no relationship between possibility and probability except for the fact that we have defined them on the same space of outcomes. Instead of formulating a relationship between the two functions directly, it is preferable, from an operational standpoint, to formulate rational desiderata for the use of impossible events in rational decision making and then attempt to derive the mathematical relationship from the application of such desiderata. In particular, it is obvious that, as an appropriate desideratum for rational decision making, we must hold that any impossible event is irrelevant to rational decision making, since taking account of an impossible event would be tantamount to allowing imaginary scenarios to dictate the outcome of our decision process. Rational decision making necessitates ignoring consideration of any events that we regard as impossible.

**DESIDERATUM 1 (Irrelevance of impossible events):** An impossible event is irrelevant to rational decision making in that it can be excluded from consideration in the sample space.

We established from Theorem 1 that an event is irrelevant to rational decision making and can be excluded from our sample space if and only if it is insignificant. Thus the above desideratum implies the following axiom relating possibility and probability.

**Axiom (Axiom of correspondence):** Any impossible event is insignificant —that is, $\nabla(\mathcal{E}) = 0$ implies that $\mathbb{P}(\mathcal{E}) = 0$.

The axiom of correspondence encapsulates the idea that the possibility and probability beliefs of the subject should be consistent —that there cannot be an impossible event with positive probability. This principle forms the first of a set of three principles that can be used to relate probability measures and possibility functions more generally in multi-valued possibility theory (see Lowen and Roubens (eds) 1993, pp. 103-112). However, for our purposes, we will deal only with all-or-nothing possibility and will therefore only require the use of this single principle.

Having obtained some relationship between the functions $\nabla$ and $\mathbb{P}$ we are now in a position to derive wider results about the relationship of possibility and probability. However, since our formulation of the notion of possibility was conducted completely independently of our formulation of probability (albeit on the same sample space of outcomes) we did not assume that $\mathcal{W} \in \mathfrak{G}$. If we assume that $\mathcal{W} \in \mathfrak{G}$ then the axiom of correspondence reduces to the following (more useful) relationship.

**Axiom (Axiom of correspondence):** This alternative statement of the axiom assumes that $\mathcal{W} \in \mathfrak{G}$. The possibility space is almost sure — that is, $\mathbb{P}(\mathcal{W}) = 1$.

The axiom of correspondence relates possibility and probability mathematically within our sample space of events. From this axiom we are able to derive further relationships between possibility, certainty and probability, which accord with our intuitive knowledge of the relationship.

**Theorem 3 (Correspondence between possibility and probability):** Within the probability space $(\mathcal{S}, \mathfrak{G}, \mathbb{P})$ and given the possibility space $\mathcal{W} \in \mathfrak{G}$ the axiom of correspondence implies that:

(a) all certain events are almost sure; and

(b) all significant events are possible.

Theorem 3 shows us the relationship between the possibility and probability under the axiom of correspondence. It shows us that all certain events are almost sure and that all significant events are possible. However, the converse is generally not true. That is, within most probability spaces there are almost sure events that are uncertain and possible events that are insignificant. This occurs when the probability measure operates within a sample space composed of a partition of possible events that is so large as to be uncountable.

**THEOREM 4 (Uncountable partitions of the sample space):** If $\mathbb{P}$ is a probability measure defined over an uncountable number of events in a partition $\mathfrak{C} \subseteq \mathfrak{G}$ then it will give zero probability to all but a countable number of these events.

**COROLLARY:** If $\mathbb{P}$ is a probability measure defined over an uncountable number of possible events in a partition $\mathfrak{C}$ then there will be an infinite number of possible but insignificant events.

In the common situation in which we are dealing with an unknown real quantity, the possibility space will often consist of an interval of real numbers (perhaps the whole real line). In this case it is possible to define a probability measure on the set of all Borel sets on this interval. This includes the singleton sets of points in the interval, which are also Borel sets, and are therefore also ascribed a probability value by the probability measure. If the interval is non-degenerate (i.e., it is not just a single point) then the singleton sets form an uncountable partition of the sample space. In this case, Theorem 4 and its corollary tell us that there will be an infinite number of singleton sets which are possible but insignificant events.

Theorem 4 and its corollary resolve one difficulty that is often faced in explaining the relationship of possibility and probability: they show why it is sometimes the case that possible but insignificant events arise in probability problems. These kinds of events arise from the extension of probability methods to deal with sets of events that are so large as to be uncountable. In fact, we can go further than this, and find conditions under which this can be avoided. Although it is not generally true that possibility and positive probability are equivalent, it turns out that they can be made to be equivalent through refinement of the probability space in certain cases. To see when this is the case, we introduce the following theorem.

**THEOREM 5 (Countable possibility space):** Within the probability space $(\mathcal{S}, \mathfrak{G}, \mathbb{P})$ and given the possibility space $\mathcal{W} \in \mathfrak{G}$, if there is a countable set of events that generates the set of significant events (this is satisfied if $\mathcal{W}$ is countable) then there is some reduction of the probability space within which the axiom of correspondence implies that:

(a) all possible events are significant; and

(b) all almost sure events are certain.

Theorem 5 shows us that if the possibility space is countable (or if a weaker condition on the set of significant events holds) then we can refine our analysis to work within a probability space in which all possible events are significant and all significant events are possible. Within such a refined space, our probabilistic beliefs are identical for all events under consideration (i.e., those events that are still under consideration after the refinement) but a sufficient number of inconsequential outcomes in the sample space have been removed to ensure that possibility and significance are identical.

While this is an interesting result, it will generally be awkward and unnecessary to work within such a probability space for the purposes of probabilistic analysis and decision making. Indeed, as we have seen, we need not be concerned with whether insignificant events are possible or not for the purposes of rational decision making.

## 7. Why possible events with zero probability arise

While the above results may be useful as part of a general exposition of the axioms of probability theory, the main purpose of this exposition has been to show that possible events with probability zero (i.e., possible but insignificant events) arise in probability theory, and to show why this is the case. We can see from the above exposition that these events arise because of the application of probability to uncountable spaces of possible outcomes.

The nature of finite measures, such as probability measures, is such that if they are defined over an uncountable partition then they must give zero measure to all but a countable number of events in the partition. If the possibility space is composed of an uncountable number of "possible states of the world" then all but a countable number of the singleton events for these states must have zero probability (assuming they are within the relevant sigma-field) —that is, there are an infinite number of possible but insignificant events.

In cases where all significant events can be generated by countable set operations on a countable set of events, the probability space can be refined to ensure correspondence between possibility and probability. However, in many cases undertaken in standard probability analysis this is not possible. In such cases there will be an infinite number of possible but insignificant events which cannot be removed from the analysis.

It is important to remember that, for the purposes of this analysis, both probability and possibility refer to epistemic concepts. This means that it is not legitimate to infer some ontological "possibilities" from epistemic probabilities, notwithstanding the reasonableness of the axiom of correspondence in the epistemic case. Just as Dubois and Prade (1993) warn against conflating degrees of probability with membership grades (p. 1060), so we warn more generally against conflating epistemological and ontological questions when assessing possibility and probability.

It is also worth noting that these mathematical results cannot be used as a substitute for philosophical argument, and so the philosophical coherence of different kinds of propositions must always be borne in mind in comparing their propositional calculus. In the words of Saul Kripke, "It should not be supposed that the formalism [of logical techniques] can grind out philosophical results in a manner beyond the capacity of ordinary philosophical reasoning. There is no mathematical substitute for philosophy" (see Evans and McDowell 1976, p. 416).

## 8. An example of possibility and probability in the multinomial model

Examples of the distinction between possibility and probability are ubiquitous, and can easily be found in any problem involving an unknown quantity with a continuous probability distribution function. As we saw from Theorem 4, examples of this kind arise any time the probability measure is defined over an uncountable partition of possible events.

However, while the theorems presented above are useful to understand the distinction, they do not really illustrate where considerations of possibility come into statistical analysis and why they are generally ignored in this context. The distinction between possibility and probability can be illustrated more effectively by considering a simple example involving a standard experimental setup in which a researcher performs some experimental action repeatedly and obtains outcomes from these actions under identical conditions during each repetition. It is common to imagine that such an experiment can be repeated indefinitely (to obtain a countably infinite sequence of outcomes) or can at least be embedded in a hypothetical repetition of this kind.

**MODEL 1 (Repetitive experiment):** Suppose we let $\boldsymbol{x} \equiv (x_1, x_2, x_3, \dots)$ be a sequence of numbers each with finite range $\{1,2,\dots,m\}$ and we let $\boldsymbol{x}_k \equiv (x_1, x_2, \dots, x_k)$. We define the *count vector* $\boldsymbol{n}_k \equiv \boldsymbol{n}_k(\boldsymbol{x}_k) \equiv (n_{k,1}, n_{k,2}, \dots, n_{k,m})$ by:

$$n_{k,a} \equiv n_{k,a}(\boldsymbol{x}_k) \equiv \sum_{i=1}^{k} \mathbb{I}(x_i = a).$$

We also define the *long-run proportion vector* $\boldsymbol{\theta} \equiv \boldsymbol{\theta}(\boldsymbol{x}) \equiv (\theta_1, \theta_2, \dots, \theta_m)$ by:

$$\theta_a \equiv \theta_a(\boldsymbol{x}) \equiv \lim_{k\to\infty} \frac{n_{k,a}}{k}.$$

One issue that may arise in such an experiment is to make a determination of the possible outcomes of the experiment under consideration. For example, in a coin-tossing experiment, there may be some question over whether it is possible for the coin to land and remain precariously balanced on its edge, without falling to either side. If this is believed to be impossible then we would be satisfied with a model that specifies only two possible outcomes for each toss: heads or tails. However, if this is believed to be possible then we would include this third possibility for each outcome.

It may also be the case that we believe that the long-run proportion of this outcome is zero. This is a weaker belief than the belief that the outcome is impossible. It requires only that we believe that the outcome will be sufficiently rare as to ensure that the long-run proportion of its occurrence will approach zero in the limiting sense (this would actually entail a continual decrease in the frequency with which it occurs over the trials). To see some of the possibility and probability beliefs we might hold, and how they relate to one another, we will consider a situation in which we want to determine whether some outcome $a$ should be included in our model.

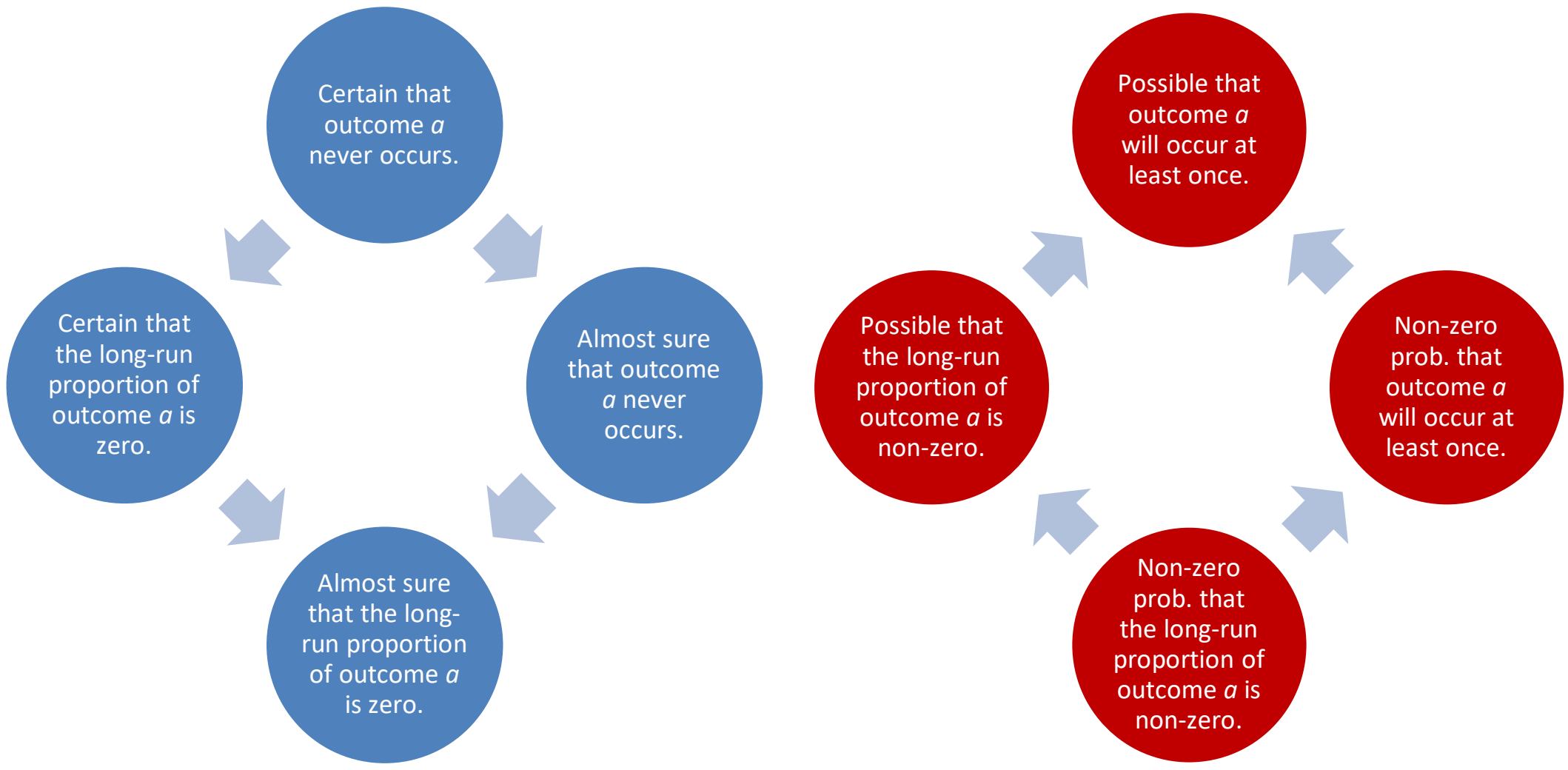

**FIGURE 1:** Implications of probability and possibility beliefs regarding outcome $a$.

Figure 1 illustrates two sets of probability and possibility beliefs with their logical implications. Each belief in the blue diagram is the opposite of the corresponding belief in the red diagram and is therefore incompatible with that belief. The overall possibility and probability belief is formed by choosing one belief from each of the corresponding colour pairs in such a way that all the logical implications are satisfied.

While these possibility and probability beliefs are interesting, one of the things we have seen from our analysis is that probability theory and decision theory based on this analysis are not concerned with whether insignificant events are possible or not. One standard probability model which arises from repeated trial experiments is the multinomial model which occurs when our probabilistic beliefs about the observable sequence are invariant to permutations in the order of the observed values. This belief, called *exchangeability*, leads us to a standard multinomial probability model and a simplification of our probabilistic beliefs about outcome $a$.

**Model 2 (Multinomial model):** In Model 1, if the sequence $\boldsymbol{x}$ is exchangeable then the elements of $\boldsymbol{x}|\boldsymbol{\theta}$ are independent with $\Pr(x_i = a|\boldsymbol{\theta}) = \theta_a$ for all $a = 1,2,\dots,m$ and $i \in \mathbb{N}$. We therefore obtain the standard multinomial probability model with:

$$\Pr(\boldsymbol{n}_k|\boldsymbol{\theta}) = \text{Mu}(\boldsymbol{n}_k|k,\boldsymbol{\theta}) = \binom{k}{\boldsymbol{n}_k}\prod_{i=1}^{m}\theta_i^{n_i}.$$

Under the multinomial model we have:

$$\Pr(\theta_a = 0) = 1 \Leftrightarrow \Pr(x_i = a) = 0;$$

$$\Pr(\theta_a > 0) > 0 \Leftrightarrow \Pr(x_i = a) > 0.$$

Moreover, since exchangeability implies a probabilistic constancy in the rate at which outcomes occur, this precludes the possibility of a possible outcome having zero long-run frequency (i.e., it ensures that all insignificant outcomes are impossible)

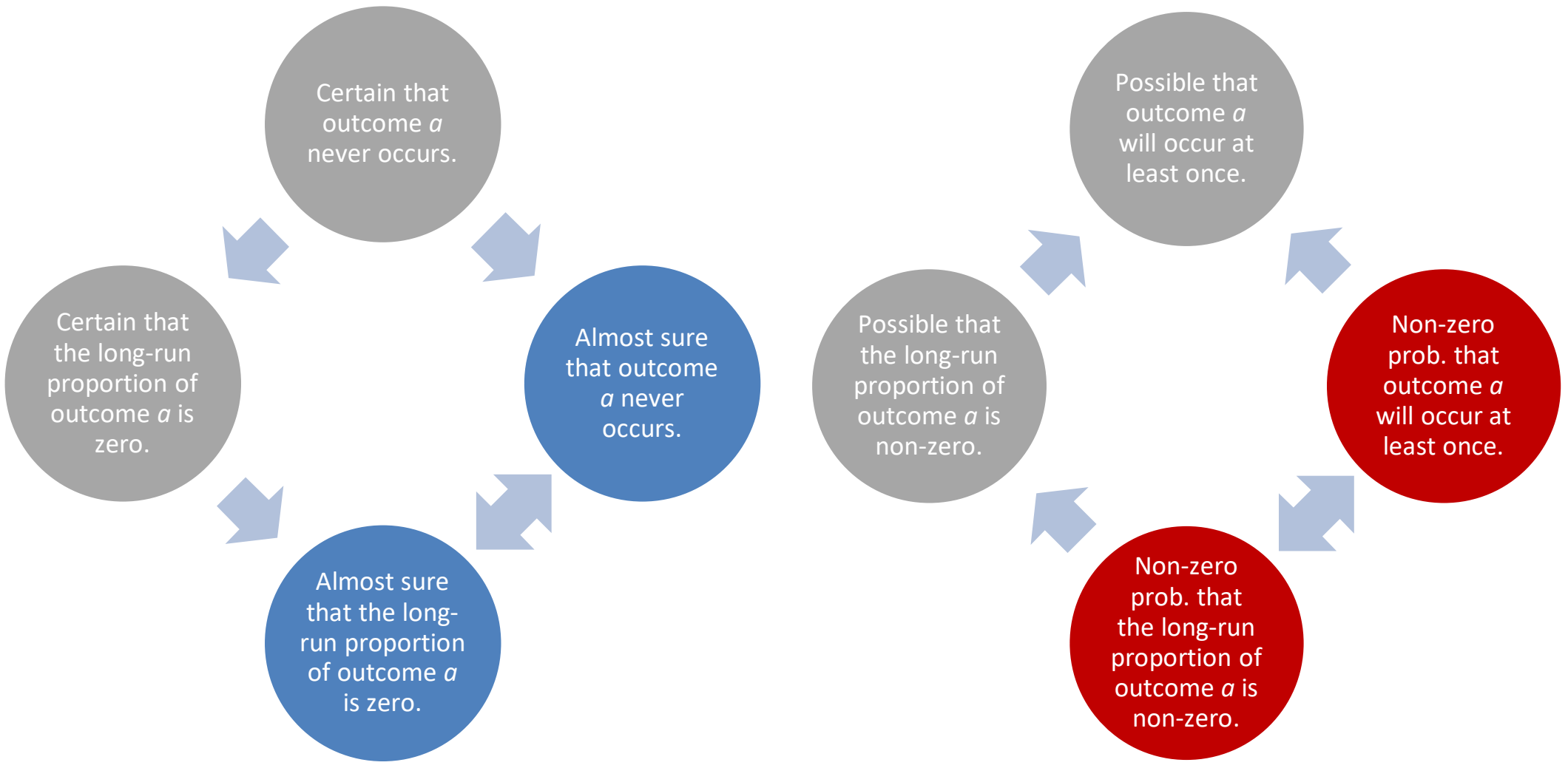

**Figure 2:** Probability beliefs about outcome *a* under the multinomial model

Figure 2 illustrates the same incompatible sets of probability and possibility beliefs as before, but now, in the context of a probabilistic analysis in the multinomial model, the only beliefs that are relevant to us are the probability beliefs. The assumption in the model leads to the logical equivalence of the two probability beliefs in either section so that the relevant considerations now boil down to just two relevant choices: either the subject is almost sure that the outcome $a$ never occurs and that the long-run proportion of this outcome is zero; or the subject ascribes some non-zero probability that outcome $a$ will occur at least once and the long-run proportion of this outcome will be non-zero.

This is not the only possibility and probability consideration that arises in the standard multinomial model. There is a further complication that arises from the definition of the long-run proportion values. Since the long-run proportion vector is determined by the sequence of outcomes, it is a function with domain given by the set of all possible sequences of the kind specified in Model 1. Unfortunately, it turns out that there are some rather nasty sequences in this set for which the limit defined in Model 1 (called the Cesàro limit) does not exist. Thus, if we want to define the long-run proportion vector on this entire domain we need to extend our definition to the Banach limit of the sequence in cases where the sequence is not Cesàro summable (the Hahn-Banach theorem ensures that the Banach limit exists in this case) (see Morrison 2000).

Unfortunately, the Banach limit is a bit more difficult to work with than the Cesàro limit since it is defined non-constructively. Thus, it would be nice to deal only with the former if we can do so without any loss of explanatory power in our analysis. To do this, we need to be able to ignore all sequences which are not Cesàro summable and thereby focus only on those sequences which have a well-defined Cesàro limit. If the probability measure for the long-run proportion vector is absolutely-continuous (with respect to Lebesgue measure) then the probability of obtaining such a sequence is one —such a sequence is not *certain*, but the event composed of all such sequences is almost sure. This allows us to create a refinement of the probability space which includes only those sequences that have a Cesàro limit. We are justified in doing so by the fact that probabilistic analysis can be refined to ignore insignificant events, even if they are possible events.

**9. Conditional probability and possibility**

The definitions and consequent properties of conditional probability are well-known to statistical practitioners and there is a large literature dealing with problems of conditional probability. It is also possible to extend our mathematical description of possibility to handle knowledge of conditioning events. This is done by observing that new knowledge of some conditioning event $E$ means that outcomes not in $E$ are now known to be false —i.e. all outcomes outside this conditioning event are now regarded as impossible. If we condition on knowledge of some conditioning event, then our possibility space is reduced by intersection with the conditioning event.

**DEFINITION 7 (Conditional possibility space and possibility function):** Given an initial possibility space $\mathcal{W} \in \mathfrak{G}$ and conditioning event $\mathcal{C} \in \mathfrak{G}$, define the conditional possibility function $\nabla: \mathfrak{G} \times \mathfrak{G} \rightarrow \{0,1\}$ by $\nabla(\mathcal{E}|\mathcal{C}) \equiv \mathbb{I}(\mathcal{E} \cap \mathcal{W} \cap \mathcal{C} \neq \emptyset)$ for all $\mathcal{E} \in \mathfrak{G}$. (We have used generic notation for simplicity.)

This conditioning relationship is very simple in the context of the standard all-or-nothing interpretation of possibility considered here. Although it can be extended to the multi-valued case of possibility functions in possibility theory, this brings with it some complications (see Dubois 2000, pp. 51-52) that we avoid here.

It can easily be shown that the same basic relationship between possibility and probability carries over from the marginal to the conditional case, regardless of the conditioning event used. That is, under any given conditioning information, it will still be the case that all certain events are almost sure and all significant events are possible. Moreover, if the conditional probability space generated by this event is such that there is a countable partition that generates the sigma-field of all significant events, then it will be possible to operate within a refined probability space where all possible events are significant and all almost sure events are certain.

We noted above that our concept of possibility was taken with respect to the actual knowledge of the individual for whom our probability space is formulated. However, it is interesting to note that our notion of conditional possibility gives us the necessary tools to extend our analysis to consider the properties of concepts of possibility pertaining to the aggregated knowledge of some other "relevant community" or knowledge not yet held but obtainable by "practicable investigation". This would allow us to extend our analysis to encompass alternative definitions of possibility put forward in works such as Hacking (1967).

Of course, it is easy to see that if we try to compare personal epistemic probabilities of the Bayesian variety with possibilities of the Hacking variety, then this must involve comparisons of probabilities and possibilities with different conditioning information.

The result is that no logical implications result, and the analysis will essentially be vacuous (though no longer ill-defined).

To see this, consider a person with beliefs (prior to any conditioning information) given by some possibility space $\mathcal{W}$ and some probability measure $\mathbb{P}$. Consider some conditioning event $\mathcal{C} \in \mathfrak{G}$ which represents either: (a) knowledge held by some other people in the "relevant community"; or (b) knowledge which could be obtained by this person through some "practicable investigation". Suppose we define possibility conditional on this additional knowledge as suggested in Hacking, while retaining the personal epistemic probability measure. In this case for any event $\mathcal{E} \in \mathfrak{G}$ we have a conditional possibility value $\nabla(\mathcal{E}|\mathcal{C})$ and a probability value $\mathbb{P}(\mathcal{E})$.

Since the former incorporates the conditioning information while the latter does not, it is easy to see that impossibility does not imply insignificance with this interpretation. In fact, if $\mathcal{E} \subseteq \mathcal{W}$ and $\mathcal{C} \subseteq \mathcal{W}$ are mutually exclusive events then $\nabla(\mathcal{E}|\mathcal{C}) = 0$ so that *regardless of its probability* $\mathcal{E}$ is an impossible event —not in the personal epistemic sense expounded in this paper, but in the sense described in Hacking.

Of course, we could certainly alter the definition of probability to accord with the Hacking interpretations of possibility by using this conditioning information in the probability measure as well as the possibility function. In this case the two concepts would again accord well with one another and we would be able to relate them sensibly by the axiom of correspondence. However, a major drawback of this approach is that it would require a rewriting of probability theory to automatically and implicitly incorporate conditioning information that does not belong to the subject under consideration. Such considerations lend credibility to the present approach of using an identical philosophical framework for both concepts, and to choosing the personal epistemic interpretation in both cases. This ensures that possibilities and probabilities mesh in accordance with our intuitive ideas and can be formally related via the axiom of correspondence.

## 10. Concluding remarks

The present paper has shown how a standard all-or-nothing conception of possibility, based on the set of possible states of the world commonly used in modal logic, can be related to probability by means of a simple axiom built on an intuitive desideratum. We have seen why possible events with zero probability arise in this paradigm and how they can (sometimes) be removed from consideration. We have also seen how possibility and probability arise in a simple model of repeated trials of an experiment, and how probability theory allows us to ignore certain complicating considerations that would be engendered by a focus on possible but insignificant events.

The theoretical views in this paper are special cases of the more general relationship between probability measures and the multi-valued possibility functions that are used in possibility theory. The utility of the present approach, confined to an all-or-nothing conception of possibility, is that it allows for a more didactic presentation of this important relationship and gives a clear connection with standard modal logic.

**Acknowledgements:** The author would like to thank two anonymous referees at the journal *Theory and Decision* where this paper was previously submitted. These referees provided valuable advice on an earlier draft of this paper.

## Appendix: Proof of Theorems

**PROOF OF THEOREM 1:** Follows easily from that fact that $\mathbb{P}_0(\mathcal{E}) = \mathbb{P}(\mathcal{E})/\mathbb{P}(\mathcal{S}_0)$ for all $\mathcal{E} \in \mathfrak{G}_0$. ■

**PROOF OF THEOREM 2(a):** We first show that an event $\mathcal{E} \in \mathfrak{G}$ is possible if and only if $\nabla(\mathcal{E}) = 1$ and is impossible if and only if $\nabla(\mathcal{E}) = 0$.

($\Longrightarrow$) If $\mathcal{E}$ is possible then it must contain at least one possible outcome so we have $\mathcal{E} \cap \mathcal{W} \neq \emptyset$, which gives $\nabla(\mathcal{E}) = \mathbb{I}(\mathcal{E} \cap \mathcal{W} \neq \emptyset) = 1$.

($\Longleftarrow$) If $\nabla(\mathcal{E}) = 1$ then we have $1 = \nabla(\mathcal{E}) = \mathbb{I}(\mathcal{E} \cap \mathcal{W} \neq \emptyset)$ giving $\mathcal{E} \cap \mathcal{W} \neq \emptyset$. This means that $\mathcal{E}$ is disjoint from the possibility space, so it contains no possible outcomes, which means it is an impossible event.

**PROOF OF THEOREM 2(b):** We now show that an event $\mathcal{E} \in \mathfrak{G}$ is certain if and only if $\nabla(\mathcal{S} - \mathcal{E}) = 0$ and is uncertain if and only if $\nabla(\mathcal{S} - \mathcal{E}) = 1$.

($\Longrightarrow$) If $\mathcal{E}$ is certain then it contains all possible outcomes so $(\mathcal{S} - \mathcal{E}) \cap \mathcal{W} = \emptyset$, which gives $\nabla(\mathcal{S} - \mathcal{E}) = \mathbb{I}((\mathcal{S} - \mathcal{E}) \cap \mathcal{W} \neq \emptyset) = 0$.

($\Longleftarrow$) If $\nabla(\mathcal{S} - \mathcal{E}) = 0$ then we have $0 = \nabla(\mathcal{S} - \mathcal{E}) = \mathbb{I}((\mathcal{S} - \mathcal{E}) \cap \mathcal{W} \neq \emptyset)$ giving $(\mathcal{S} - \mathcal{E}) \cap \mathcal{W} = \emptyset$. Since $\mathcal{S} \supseteq \mathcal{W}$ this implies that $\mathcal{E} \supseteq \mathcal{W}$ which means that $\mathcal{E}$ contains all possible outcomes, so it is a certain event.

**PROOF OF THEOREM 3:** (a) If $\mathcal{E}$ is certain then $\mathcal{S} - \mathcal{E}$ is impossible. It follows from the axiom of correspondence that $\mathcal{S} - \mathcal{E}$ is insignificant so that $\mathcal{E}$ is almost sure. (b) Follows directly from the fact that all impossible events are insignificant. ■

**PROOF OF THEOREM 4:** Let $\mathfrak{C}_+ \equiv \{\mathcal{E} \in \mathfrak{C}: \mathbb{P}(\mathcal{E}) > 0\}$ be the set of all events in the partition that are significant. The theorem requires us to show that $\mathfrak{C}_+$ is countable. To do this, we decompose $\mathfrak{C}_+ = \bigcup_{k=1}^{\infty} \mathfrak{U}_k$ where the sets $\mathfrak{U}_1, \mathfrak{U}_2, \mathfrak{U}_3, \ldots$ are defined by:

$$\mathfrak{U}_k \equiv \left\{\mathcal{E} \in \mathfrak{C}: \frac{1}{k+1} < P(\mathcal{E}) \leq \frac{1}{k}\right\} \quad \text{for all } k \in \mathbb{N}.$$

Now, suppose —contrary to the theorem— that $\mathfrak{C}_+$ is uncountable. This implies that at least one of the sets $\mathfrak{U}_1, \mathfrak{U}_2, \mathfrak{U}_3, \ldots$ must be an infinite set. Suppose, without loss of generality that $\mathfrak{U}_h$ is an infinite set for some $h \in \mathbb{N}$. Since $\mathfrak{C}$ is a partition, all the events in $\mathfrak{C}$ are disjoint, which means that all the events in $\mathfrak{U}_h$ are disjoint. Thus, we can choose a countable sequence of disjoint events $\mathcal{E}_1, \mathcal{E}_2, \mathcal{E}_3, \ldots \in \mathfrak{U}_h$ and obtain:

$$\mathbb{P}\left(\bigcup_{i=1}^{\infty} \mathcal{E}_i\right) = \sum_{i=1}^{\infty} \mathbb{P}(\mathcal{E}_i) \geq \sum_{i=1}^{\infty} \frac{1}{h+1} = \infty.$$

This means that $\mathbb{P}$ is not a probability measure, which contradicts the premise of the theorem. Thus, by contradiction, $\mathfrak{C}_+$ must be countable, which was to be shown. ■

**Proof of Theorem 5:** Let $\mathfrak{G}_+ \equiv \{\mathcal{E} \in \mathfrak{G}: \mathbb{P}(\mathcal{E}) > 0\}$ be the set of all the events that are significant. If there is a countable set of events generating $\mathfrak{G}_+$ then there must also be a countable set $\mathfrak{C}_+ \subseteq \mathfrak{G}_+$ of *disjoint* events generating $\mathfrak{G}_+$. (If $\mathcal{W}$ is countable then $\mathfrak{C}_+$ can consist of all significant singleton sets of the points in $\mathcal{W}$.) Letting $\mathcal{S}_0 \equiv \bigcup \mathfrak{G}_+ = \bigcup \mathfrak{C}_+$ we have $\mathcal{W} - \mathcal{S}_0 \in \mathfrak{G} - \mathfrak{G}_+$ so that $\mathbb{P}(\mathcal{W} - \mathcal{S}_0) = 0$. From the axiom of correspondence we have $\mathbb{P}(\mathcal{W}) = 1$ so that $\mathbb{P}(\mathcal{S}_0) = 1$. We therefore obtain a reduction $(\mathcal{S}_0, \mathfrak{G}_0, \mathbb{P}_0)$ by conditioning on the almost sure event $\mathcal{S}_0$ giving us a sigma-field of events $\mathfrak{G}_0$ being generated by $\mathfrak{C}_+$. If $\mathcal{E} \in \mathfrak{G}_0$ then $\mathcal{E} \supseteq \mathcal{E}_+$ for some $\mathcal{E}_+ \in \mathfrak{C}_+$ so that:

$$\mathbb{P}(\mathcal{E}) \geq \mathbb{P}(\mathcal{E}_+) > 0.$$

Thus, within this refined space, all events are possible and significant. Since this is the case, it follows that the whole space is the only almost sure event in the refined sigma-field of events and this event is also certain. ■